\documentclass[12pt]{article}

\usepackage[dvips]{graphicx}
\usepackage[english]{babel}

\usepackage{amssymb,epsfig}
\usepackage{amsmath}

\usepackage{slashed}
\usepackage[active]{srcltx}
\usepackage{psfrag}

\newcommand{\ms}{\mskip 1.5mu}

\begin{document}

\begin{flushright}
%draft of \today
SI-HEP-2012-19 
\end{flushright}

\bigskip

\begin{center}
\textbf{\LARGE  Soft contribution to $B\to \gamma \ell \nu_\ell$ and 
the $B$-meson distribution amplitude}

\vspace*{1.6cm}

{\large V.M. Braun$\ms{}^1$ and  A.~Khodjamirian$\ms{}^{2}$}
 
\vspace*{0.4cm}

\textsl{%
$^1$ Institut f\"ur Theoretische Physik, Universit\"at Regensburg,
D-93040 Regensburg, Germany \\
$^2$ Theoretische Physik 1,
Naturwissenschaftlich-Technische Fakult\"at,\\
Universit\"at Siegen, D-57068 Siegen, Germany}

\vspace*{0.8cm}

\textbf{Abstract}\\[10pt]
\parbox[t]{0.9\textwidth}{
The $B\to \gamma \ell \nu_\ell$  decay at large energies of the photon 
receives a numerically important soft-overlap contribution which is 
formally of the next-to-leading order in the expansion in the  
inverse photon energy. 
We point out that this contribution can be calculated 
within the framework of heavy-quark expansion  and soft-collinear effective theory, 
making use of dispersion relations and quark-hadron duality. The soft-overlap contribution 
is obtained in a full analogy with 
the similar contribution to the $\gamma^* \gamma \to \pi$ transition form factor.  
This result strengthens the case for using the $B\to \gamma \ell \nu_\ell$ 
decay to constrain the $B$-meson distribution 
amplitude and determine its most important parameter, the inverse moment $\lambda_B$.  
}
\end{center}

\vspace*{1cm}

%%%%%%%%%%%%%%%%%%%%%%%%%%%%%%%%%%%%%%%%%%%%%%%%%%

%\section{Introduction}
{\large \bf 1.}~~ The  decay $B\to \gamma \ell \nu_\ell$  
at  large photon energy $E_\gamma$  is one of the simplest hadronic processes that can be studied 
using QCD factorization~\cite{Korchemsky:1999qb,DescotesGenon:2002mw,Lunghi:2002ju,Bosch:2003fc}. This method involves the $B$-meson light-cone distribution amplitude (DA)~\cite{Grozin:1996pq,Beneke:2000wa,Kawamura:2001jm,Braun:2003wx,Lee:2005gza,Kawamura:2008vq} 
as the main nonperturbative input at the leading order in the heavy quark expansion.
The $B\to \gamma\ell \nu_\ell$  decay \footnote{to distinguish this important 
decay channel we suggest to call it {\em photoleptonic} $B$-decay.} is therefore best suited for determining the 
parameters of the $B$-meson DA  and presents a close analog 
to the photon-pion transition $\gamma^* \gamma\to \pi$ with one 
highly-virtual ($Q^2$) and one real photon, which  is used for the determination of the pion DA.
The theoretical challenges are also similar, and in both cases are mainly due to the 
necessity to have a quantitative control
over the corrections that are subleading in the expansion parameter, 
$1/m_b\sim 1/(2E_\gamma)$ or $1/Q^2$, and, in general, are not factorizable.
The purpose of this letter is to emphasize this connection and suggest a method 
to calculate the soft contribution to  
$B\to \gamma \ell \nu_\ell$, which closely follows the technique used in $\gamma^* \gamma\to\pi$. 

A recent summary of the theory status of $B\to \gamma \ell \nu_\ell$ can be found in \cite{Beneke:2011nf}.
The decay amplitude
\begin{equation}
A(B^-\to \gamma \ell \bar{\nu}_\ell)=\frac{G_FV_{ub}}{\sqrt{2}}
\langle \ell\bar{\nu}_l\gamma | \bar{\ell}\,\gamma^\nu(1-\gamma_5)\nu_\ell
\bar{u}\gamma_\nu(1-\gamma_5)b |B^-\rangle\,
\label{eq:defampl}
\end{equation}
can be written in terms of the two form factors, $F_V$ and $F_A$, defined through the Lorentz decomposition  of the hadronic tensor
\begin{eqnarray}
T_{\mu\nu}(p,q)&=&-i\int d^4x\,e^{ipx}\langle 0 | T\{ j^{em}_\mu(x)\,\bar{u}(0)\gamma_\nu(1-\gamma_5)b(0)\}| B^- (p+q)\rangle
\nonumber\\&=& 
\epsilon_{\mu\nu\tau \rho }p^\tau v^\rho  F_V +i\big[-g_{\mu\nu}(pv)+v_\mu p_\nu \big ]F_A+ \ldots\,.
\label{eq:Tmunu}
\end{eqnarray}
Here $p$ 
and $q$  are the photon and lepton-pair momenta, respectively, 
so that $(p+q)= m_B v$ is the $B$-meson momentum in terms of its velocity.
In the above,  
$j_{\rm em}^\mu=\sum_q e_q\bar{q}\gamma_\mu q$ is the electromagnetic current and the ellipses stand for the  terms $\sim p_\mu$ and for the contact term. The origin
of the latter is explained  in \cite{Beneke:2011nf} (see also \cite{Khodjamirian:2001ga}). 
The form factors can be written as functions of the lepton-pair invariant mass squared $q^2$, or, equivalently, 
of the photon energy $E_\gamma$ in the $B$-meson rest frame. The differential decay width is given by
\begin{equation}
 \frac{d\Gamma}{dE_\gamma} = 
\frac{\alpha_{\rm em}G_F^2|V_{ub}|^2}{6\pi^2}m_B E_\gamma^3\left(1-\frac{2E_\gamma}{m_B}\right)
\Big(|F_V|^2+|\widetilde{F}_A|^2\Big)\,,\qquad 
\widetilde{F}_A = F_A + \frac{e_\ell f_B}{E_\gamma}\,,
\label{eq:width}
\end{equation}
where the contact term is included in the axial form factor, and the lepton mass is neglected.

For large photon energies the form factors can be calculated \cite{Beneke:2011nf} as
\begin{eqnarray}
  F_V(E_\gamma) &=& \frac{e_u f_B m_B}{2 E_\gamma \lambda_B(\mu)} R(E_\gamma,\mu) + 
\Big[\xi(E_\gamma) + \frac{e_b f_B m_B}{2 E_\gamma m_b}+\frac{e_u f_B m_B}{(2 E_\gamma)^2}\Big],
\nonumber\\
  F_A(E_\gamma) &=& \frac{e_u f_B m_B}{2 E_\gamma \lambda_B(\mu)} R(E_\gamma,\mu) + 
\Big[\xi(E_\gamma) - \frac{e_b f_B m_B}{2 E_\gamma m_b}-\frac{e_u f_B m_B}{(2 E_\gamma)^2}\Big].
\label{eq:SCET1}
\end{eqnarray}
The first term in both expressions represents the leading contribution in the heavy-quark expansion (HQE) 
that corresponds to the photon emission from the light spectator quark in $B$ meson. 
In the above, $f_B$ is the decay constant of $B$ meson, and
the quantity $\lambda_B$  is the first inverse moment of the $B$-meson DA:
\begin{equation}
  \frac{1}{\lambda_B(\mu)} = \int\limits_0^\infty\frac{d\omega}{\omega}\phi_+(\omega,\mu)\,,
\end{equation}
where the variable $\omega$ is the energy  of the light-quark 
in the $B$-meson (in its rest frame). 
The factor 
$R(E_\gamma,\mu)$  in (\ref{eq:SCET1}) takes into account gluon radiative corrections 
(see \cite{Beneke:2011nf} for details) and equals one at the tree level. 
The scale-dependence of $\phi_+(\omega,\mu)$ was calculated to the leading-logarithmic accuracy in \cite{Lange:2003ff}.

The terms in square brackets in (\ref{eq:SCET1}) are the $1/m_b$ and $1/(2E_\gamma)$ power corrections to 
the leading-order expression.
They are written in this particular form to emphasize that some of these corrections are ``symmetry-preserving'',
i.e. are the same for both form factors $F_V$ and $F_A$, and some corrections are ``symmetry-breaking'', i.e. 
they are different. The symmetry-preserving corrections parametrized by the function $\xi(E_\gamma)$ 
present the main difficulty. They are unknown, apart that from 
the power counting
one expects $\xi(E_\gamma)\sim 1/E_\gamma$ with respect to the leading-order term. 
In the analysis of \cite{Beneke:2011nf} this function was parametrized as
\begin{equation}
 \xi(E_\gamma) = c \,\frac{f_B}{2 E_\gamma}\,,
\label{eq:cxi}
\end{equation} 
that is, tacitly assuming a $1/m_B$ suppression with respect to the leading-order term
in (\ref{eq:SCET1}),  and the coefficient $c$ was varied between $-1$ and $+1$. 

In this letter we present an approach which allows
one to estimate $\xi(E_\gamma)$ to, potentially, 20\% accuracy. To be precise, we will be calculating the 
soft overlap contribution to $\xi(E_\gamma)$ which is not directly accessible in QCD factorization.
Generally speaking, there exist also factorizable symmetry-preserving contributions which can treated in a 
systematic way within the HQE and added.
%We will comment on the separation between factorisable and soft-overlap contributions in what follows.         

%%%%%%%%%%%%%%%%%%%%%%%%%%%%%%%%%%%%%%%%%%%%%%%%%%%%%%%%%%
%%%%%%%%%%%%%%%%%%%%%%%%%%%%%%%%%%%%%%%%%%%%%%%%%%%%%%%%%%

\vspace*{0.3cm}

{\large \bf 2.}~~The main idea which parallels the technique originally suggested in~\cite{Khodjamirian:1997tk} 
for $\gamma^*\gamma\to \pi$ transition form factor 
is to consider the hadronic tensor (\ref{eq:Tmunu}) at negative $p^2<0$, 
$m_B^2 \gg |p^2| \gg \Lambda_{\rm QCD}^2$, or in other words, an  
unphysical decay $B\to \gamma^* \ell \nu_\ell$  involving a (transversely polarized) spacelike photon. The corresponding 
form factors, now functions of two variables $q^2$ and $p^2$, can be calculated 
in QCD using HQE and operator-product expansion (OPE) to 
(at least in principle) arbitrary power accuracy in $1/m_b,1/E_\gamma,1/p^2$.  
The analytic continuation to the real photon limit 
$p^2=0$ can be made using dispersion relation. In this way the explicit evaluation of soft nonfactorizable contributions
is effectively replaced by a certain ansatz
of the hadronic spectral density in the $p^2$-channel. 

The starting observation (cf.~\cite{Khodjamirian:1997tk}) is that $F_{V,A}(q^2,p^2)$, the generalized form factors of $B\to \gamma^* \ell \nu_\ell$, at fixed $q^2$
satisfy an unsubtracted dispersion relation in the variable $p^2$.
Separating the contribution of the lowest-lying vector mesons $\rho,\omega$, one can write ($F_{B\to \gamma^*} =F_V$ or $ F_A$)
\begin{equation}
F_{B\to \gamma^*}(q^2,p^2)= \frac{f_\rho F_{B\to \rho}(q^2)}{m^2_\rho-p^2} + 
\frac{1}{\pi}\int\limits_{s_0}^\infty ds\,\frac{\mathrm{Im} F_{B\to \gamma^*}(q^2,s)}{s-p^2}\,,
\label{eq:DR}
\end{equation}
where $s_0$ is a certain effective threshold. Here, we combined the $\rho$ and $\omega$ 
contributions  in one resonance term, assuming $m_\rho\simeq m_\omega$ and 
adopting the zero-width
approximation. In the above, $f_\rho$  is the usual decay constant
of the vector meson and $F_{B\to \rho}(q^2)$ is a generic $B\to \rho(\omega)$ 
transition form factor. 
Note that since there are no massless states, the real photon limit is recovered 
by the simple substitution $p^2\to 0$ in (\ref{eq:DR}).  

On the other hand, the same form factors can be calculated using HQE. 
The result satisfies a similar dispersion relation
\begin{equation}
 F^{\rm HQE}_{B\to \gamma^* }(q^2,p^2) = \frac{1}{\pi}\int\limits_{0}^\infty ds\,\frac{\mathrm{Im} F^{\rm HQE}_{B\to \gamma^*}(q^2,s)}{s-p^2}\,,
\label{eq:DRQCD}
\end{equation}
where the limit $p^2\to 0$ cannot be taken directly: e.g., singular terms in $1/p^2$ appear (cf.~\cite{Agaev:2010aq}), signaling that the HQE cannot be 
applied to the real photon case $p^2=0$ beyond the leading power in $1/m_b$.
 
The main assumption of the method is that the physical spectral density 
above the threshold $s_0$ coincides with the QCD spectral density, as given by the HQE:
\begin{equation}
   \mathrm{Im}F_{B\to \gamma^*}(q^2,s) \simeq \mathrm{Im}F^{HQE}_{B\to \gamma^* }(q^2,s)\qquad \mbox{for}~~s>s_0\,.
\label{eq:duality}
\end{equation}
This is the usual approximation of local quark-hadron duality.
In reality we employ a weaker form of duality, assuming that 
(\ref{eq:duality}) holds upon averaging
with a smooth weight function over a sufficiently broad interval of the 
variable $s$.

We expect that the HQE reproduces the ``true'' form factors $F_{B\to \gamma^*}(q^2,p^2)$  for 
asymptotically large values of $-p^2$. 
Equating the two representations (\ref{eq:DR}) and (\ref{eq:DRQCD}) at $p^2\to-\infty$ 
and subtracting the contributions of $s>s_0$ from both sides one obtains
\begin{eqnarray}
  f_\rho F_{B\to \rho }(q^2) = \frac{1}{\pi}\int\limits_{0}^{s_0}ds\, \mathrm{Im} F^{\rm HQE}_{B\to \gamma^* }(q^2,s)\,.
\end{eqnarray}
This relation explains why $s_0$ is usually referred to as the interval of duality 
in the vector-meson channel.
The perturbatively obtained  
spectral density in HQE 
$\mathrm{Im}F^{HQE}_{B\to \gamma^* }(q^2,s)$ is a
smooth function of $s$ and does not vanish at $s\to 0$. It is very different from the hadronic spectral 
density $\mathrm{Im}F_{B\to \gamma^* \ell \nu_\ell}(q^2,s) \sim \delta(s-m_\rho^2)$.
However, the integrals over both  
spectral densities over a certain region of $s$ 
coincide; in this sense the QCD  
description of correlation functions in terms of quark and gluons is dual to the one in terms of hadronic states. 

In practical applications of this method one uses an additional device, borrowed from the method of 
QCD sum rules \cite{SVZ}, which allows one to reduce the sensitivity to the duality
assumption in (\ref{eq:duality}) and simultaneously suppress the contributions of higher orders 
in the OPE.  This is done going over to the Borel representation $1/(s-p^2)\to \exp[-s/M^2]$,
the net effect being the appearance of an additional weight factor under the integral     
\begin{eqnarray}
  f_\rho F_{B\to \rho }(q^2) = \frac{1}{\pi}\int\limits_{0}^{s_0}ds\, e^{-(s-m^2_\rho)/M^2}\,
\mathrm{Im} F^{\rm HQE}_{B\to \gamma^* }(q^2,s)\,.
\label{eq:Frhogamma}
\end{eqnarray}
Varying the Borel parameter within a certain window, usually $M^2=1-2$~GeV$^2$, one can test the 
sensitivity of the results to the particular model of the spectral density
\footnote{We note in passing that the relation (\ref{eq:Frhogamma}) where 
the r.h.s. is calculated in terms of HQE and  $B$ meson DA's, represents a  
{\em light-sone sum rule} for $B\to \rho$ form factor of the particular type considered
in \cite{Khodjamirian:2005ea,DeFazio:2005dx}.}. 

With this refinement, substituting (\ref{eq:Frhogamma}) in (\ref{eq:DR}) and using 
(\ref{eq:duality}) one obtains for $p^2\to 0$ (cf.~\cite{Khodjamirian:1997tk,Agaev:2010aq})
\begin{equation}
  F_{B\to \gamma }(q^2) = 
\frac{1}{\pi}\int\limits_{0}^{s_0} \frac{ds}{m_\rho^2} \mathrm{Im} F^{\rm HQE}_{B\to \gamma^* }(q^2,s)e^{-(s-m^2_\rho)/M^2} +
\frac{1}{\pi}\int\limits_{s_0}^\infty \frac{ds}{s} \mathrm{Im} F^{\rm HQE}_{B\to \gamma^* }(q^2,s)\,. 
\label{eq:general}
\end{equation}
This expression contains two nonperturbative parameters --- the vector meson mass $m_\rho$ and effective threshold 
$s_0\simeq 1.5$~GeV$^2$---
as compared to the ``pure'' QCD calculation based on the HQE expansion. The reward is that the HQE can be done to an
arbitrary accuracy in the powers of $1/m_b$. The nonfactorizable contributions that are beyond the accuracy of the HQE in the usual treatment
are taken into account effectively, via the nonperturbative modification of the spectral density. 
\vspace*{0.3cm}

{\large \bf 3.}~~
As an illustration, consider the 
expression corresponding to the leading-order in HQE diagram of the virtual photon
emission from the light spectator-antiquark in $B$ meson. Calculating this diagram 
in terms of $B$-meson DA  at $p^2<0$, we obtain, after replacing 
$q^2=m_B^2-2m_BE_\gamma+p^2$,
the following expression  for the form factors defined in (\ref{eq:Tmunu}):  
\begin{equation}
F^{(0)}_V(q^2,p^2) =  F^{(0)}_{B\to \gamma^*}(E_\gamma,p^2) = e_u f_B m_B
\int\limits_0^\infty{d\omega}\frac{\phi_+(\omega,\mu)}{2 E_\gamma \omega - p^2}\,,
\label{eq:Fzero} 
\end{equation}
where we neglected the corrections $\sim \omega/E_\gamma$. In this 
approximation, the axial form factor $F^{(0)}_A(q^2,p^2)$ coincides with
$F^{(0)}_V(q^2,p^2)$.
The energy integral in (\ref{eq:Fzero}) can easily be converted to the form of a dispersion 
relation by the change of variables $s = 2 E_\gamma \omega$. 
{}Following the procedure described above and changing the integration variable back to
$\omega = s/(2E_\gamma)$, we obtain, at $p^2\to 0$, 
\begin{equation}
  F^{(0)}_{B\to \gamma }(E_\gamma) =
  \frac{e_u f_B m_B}{2E_\gamma} \biggl[
  (2 E_\gamma)\!\!\!\int\limits_0^{s_0/(2E_\gamma)}\!\! \frac{d\omega}{m_\rho^2}\,\,\phi_+(\omega,\mu)e^{-(2E_\gamma\omega- m_\rho^2)/M^2}
 + 
 \int\limits_{s_0/(2E_\gamma)}^\infty\!\! \frac{d\omega}{\omega}\,\phi_+(\omega,\mu) \biggr].
\label{eq:LCSR_LO}
\end{equation}
Completing the second integral to run from zero to infinity and subtracting the correction from the first term, we can
write this expression as
\begin{equation}
  F^{(0)}_{B\to \gamma}(E_\gamma) = \frac{e_u f_B m_B}{2E_\gamma\lambda_B(\mu)} + 
   \frac{e_u f_B m_B}{2E_\gamma}
              \!\!\!\int\limits_0^{s_0/(2E_\gamma)}\!\!d\omega \biggl[\frac{2E_\gamma}{m^2_\rho}e^{-(2E_\gamma\omega- m_\rho^2)/M^2}-\frac{1}{\omega}\biggr]
\phi_+(\omega,\mu)\,,
\label{eq:LCSR_LO1}
\end{equation}
where the first term is nothing but the HQE expression for the form factor to leading order in $\alpha_s$ and leading power accuracy in 
(\ref{eq:SCET1}). The second term can be identified with the soft correction
 $\xi(E_\gamma)$  as it appears in the same equations.
Note that the modification of the standard HQE expression only concerns the region $\omega < s_0/(2E_\gamma)\sim 
s_0/m_b$, hence, this is a soft spectator-quark contribution. 
If $\phi_+(\omega,\mu) \sim \omega $ for $\omega\to 0$%
~\cite{Grozin:1996pq,Kawamura:2001jm,Braun:2003wx,Lee:2005gza,Kawamura:2008vq}, $\xi^{(0)}(E_\gamma)$
defined by (\ref{eq:LCSR_LO1}) corresponds to a power correction of the order of 
$s_0/(2E_\gamma)^2$ for $E_\gamma\sim m_b \to \infty$, in agreement with usual power counting.

We define a rescaled soft contribution $\widehat{\xi}(E_\gamma)$ such that 
\begin{eqnarray}
 F_{B\to \gamma}(E_\gamma) & = & \left(\frac{e_u f_B m_B}{2E_\gamma\lambda_B(\mu)}\right)
\left(1+\frac{\widehat{\xi}}{2 E_\gamma}\right) +\ldots\,,
\label{hatxi}
\end{eqnarray}
where the expression in the parenthesis is the leading-order result, and we anticipate that the function $\widehat{\xi}(E_\gamma)$ depends on  $E_\gamma$ only weakly.
The ellipses stand for radiative and ``hard'' power corrections, 
cf. (\ref{eq:SCET1}) \footnote{ E.g., the corrections given by  
the $\sim e_u$ terms in square brackets in (\ref{eq:SCET1}) are readily 
obtained if one retains the $O(\omega/E_\gamma)$ terms in the integral (\ref{eq:Fzero}).}.
In the adopted approximation
\begin{equation}
    \widehat{\xi}^{(0)}(E_\gamma) = 2E_\gamma \lambda_B \int\limits_0^{s_0/(2E_\gamma)}\!\!d\omega 
\biggl[\frac{2E_\gamma}{m^2_\rho}e^{-(2E_\gamma\omega- m_\rho^2)/M^2}-\frac{1}{\omega}\biggr]
\phi_+(\omega,\mu)\,. 
\label{hatxi1}
\end{equation}
Note that the definition of the $B$-meson DA in the soft-collinear effective theory (SCET) 
involves a collinear light spectator-antiquark field.
If the separation between collinear and soft regions were done with an explicit  
cutoff, $\omega > \mu_{\rm SC}^2/E_\gamma$, the leading-order result for the form factors would read 
\begin{equation}
F^{SCET}_{B\to \gamma }(E_\gamma) =
  \frac{e_u f_B m_B}{2E_\gamma} 
 \int\limits_{\mu_{\rm SC}^2/E_\gamma}^\infty\!\! \frac{d\omega}{\omega}\,\phi_+(\omega,\mu)\,.
\label{SCET2}
\end{equation} 
It is easy to see that the soft correction defined by
(\ref{eq:LCSR_LO}), (\ref{eq:LCSR_LO1}) effectively cuts off the small energy region 
in a way similar in spirit to (\ref{SCET2}), with the interval of duality 
playing the role of the hard cutoff $\mu_{\rm SC}$. 
We therefore expect that the soft nonperturbative correction 
is always \emph{negative} relative to the leading-order result because its role is,
conceptually, to create a mass gap in the vector-meson mass spectrum.  

%%%%%%%%%%%%%%%%%%%%%%%%%%%%%%%%%%%%%%%%%%%%%%%%%%%%%%%%%%
%%%%%%%%%%%%%%%%%%%%%%%%%%%%%%%%%%%%%%%%%%%%%%%%%%%%%%%%%%

\vspace*{0.3cm}

{\large \bf 4.}~~
For numerical estimates we will use two models of the $B$-meson DA:
\begin{eqnarray}
 \phi^{I}_+(\omega)&=&\frac{\omega}{\lambda_B^2}e^{-\omega/\lambda_B} \qquad \cite{Grozin:1996pq}\,,
\nonumber\\
\phi^{II}_+(\omega)&=& \frac{4}{\pi\lambda_B}\frac{\hat{\omega}}{\hat{\omega}^2+1}
                  \Big[\frac{1}{\hat{\omega}^2+1}- \frac{2(\sigma_B-1)}{\pi^2}\ln \hat{\omega} \Big]\,,\qquad \cite{Braun:2003wx}\,, 
\label{eq:phi+}
\end{eqnarray}
where $\hat{\omega}= \omega/1$~GeV and $\sigma_B$ is the first logarithmic moment
%\begin{equation}
  $\sigma_B \lambda_B^{-1} \!= \!-\!\int\limits_0^\infty \frac{d\omega}{\omega}\,\phi_+(\omega,\mu) \ln\frac{\omega}{\mu}$.
%\end{equation}
We take $300\, \text{MeV} < \lambda_B < 600 $~MeV and $\sigma_B = 1.5$~\cite{Braun:2003wx,Lee:2005gza} 
(see also \cite{Kawamura:2008vq}) as typical values of the parameters.

\begin{figure}[t]
\psfrag{x}[cc][cc][0.9]{$2 E_\gamma\,\text{[GeV]}$}
\psfrag{y}[cc][cc][0.9]{$\widehat{\xi}_{B\to \gamma \ell \nu_\ell}\,\text{[GeV]}$}
\psfrag{a}[cc][cc][0.9]{$Q^2\,[\text{GeV}^2]$}
\psfrag{b}[cc][cc][0.9]{$\widehat{\xi}_{\gamma^*\gamma\to\pi}\,[\text{GeV}^2]$}
\begin{center}
\includegraphics[width=.45\textwidth,clip=true]{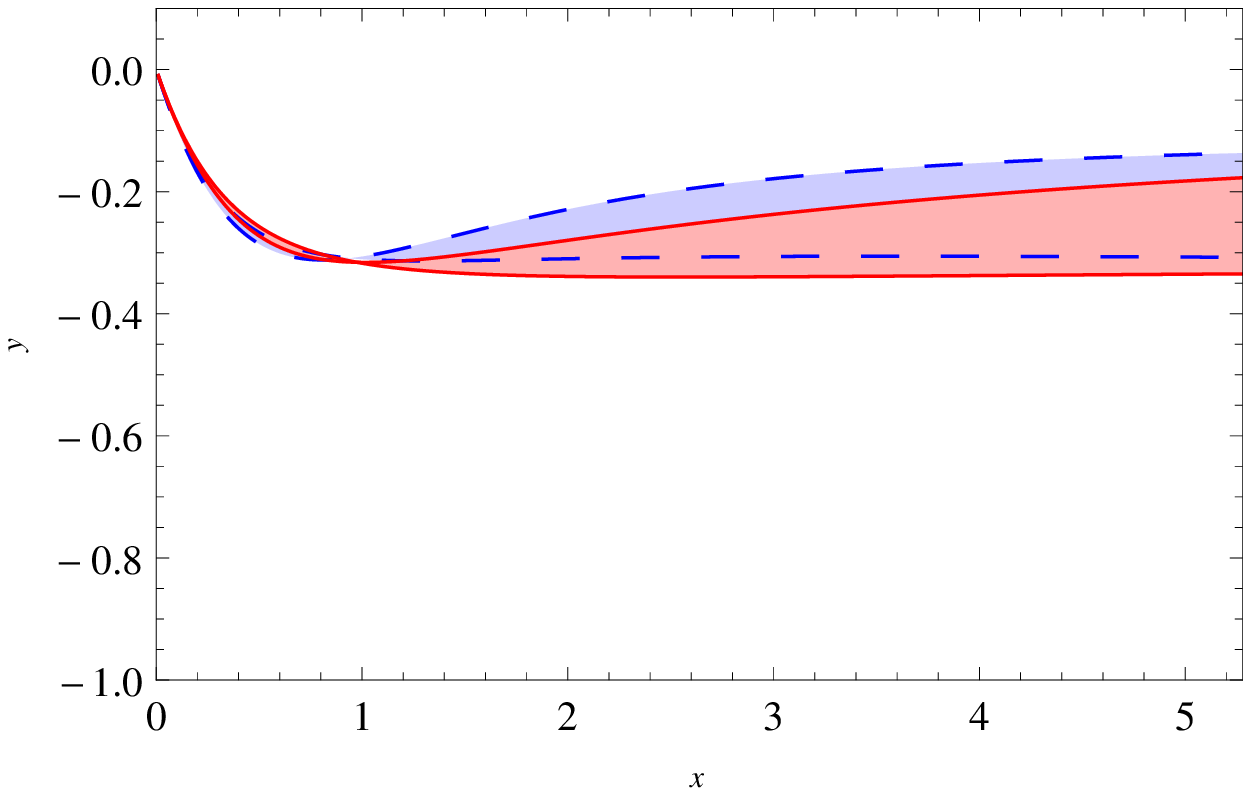}\hspace*{0.6cm}
\includegraphics[width=.45\textwidth,clip=true]{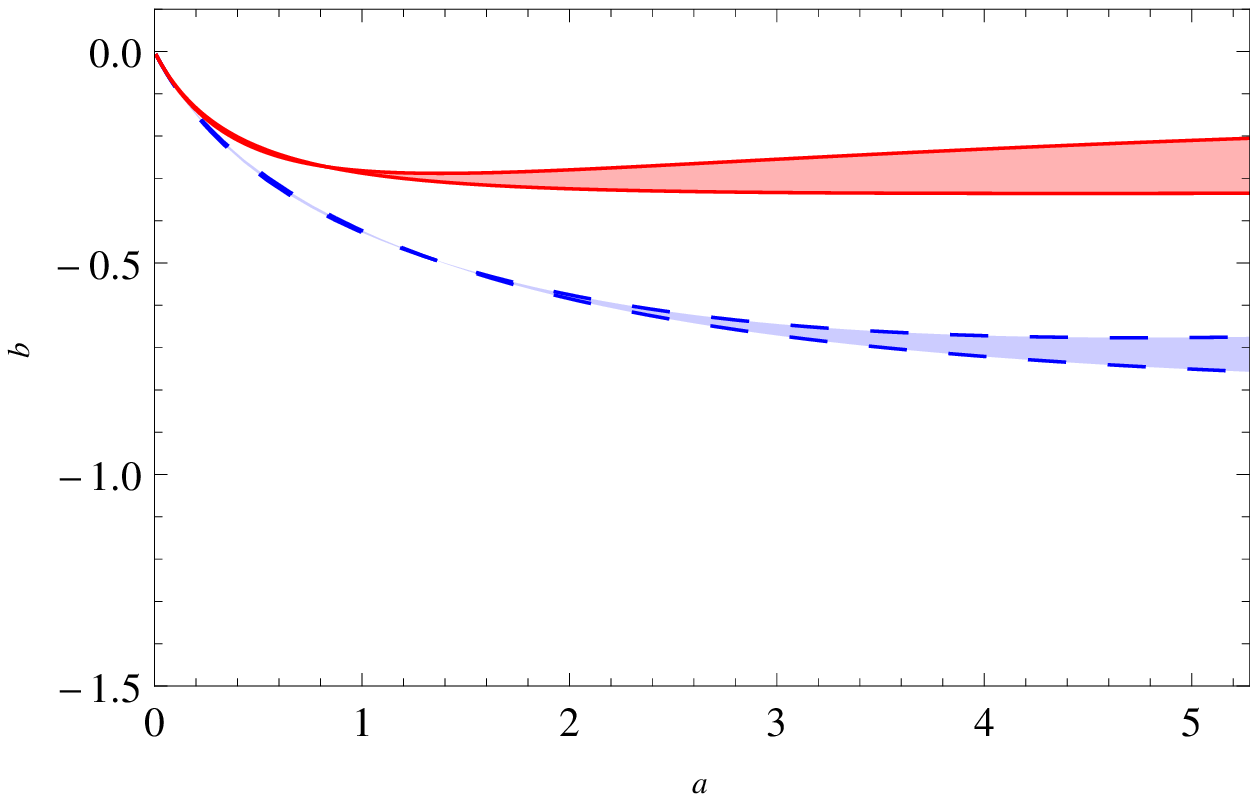}
\end{center}
\caption{\small\sf 
The soft contribution $\widehat{\xi}(E_\gamma)$ to the form factors 
in $B\to \gamma \ell \nu_\ell$ 
(left panel) for the first (solid curves) and second (dashed curves) model of the $B$-meson DA in  (\ref{eq:phi+}),
compared to the soft contribution $\widehat{\xi}(Q^2)$ to the $\gamma^*\gamma\to\pi$ 
form factor (right panel) for the asymptotic pion DA (solid) and a realistic model with 
$a_{2,4}\neq 0$ (dashed). The shaded areas correspond to the variation of the Borel parameter in the range  
$M^2=1.0- 1.5$~GeV$^2$.
}
\label{fig:xihat}
\end{figure}

The function $\widehat{\xi}(E_\gamma)$ calculated for the two models of the $B$-meson DA 
assuming the standard choice of the continuum threshold 
$s_0=1.5$~GeV$^2$ and $M^2=1.0-1.5$~GeV$^2$ is plotted vs. $2 E_\gamma$ in 
Fig.~\ref{fig:xihat}, the left panel.
For this plot we have chosen $\lambda_B= 500$~MeV. 
Note that the curves are essentially flat for $2E_\gamma > 1$~GeV which means that the soft 
contribution is well described by a power-suppressed correction $\sim 1/(2E_\gamma)$ as compared to the 
leading-order result. 

For comparison, we show in the same figure (right panel) the soft correction to the 
$\gamma^*\gamma\to \pi$ transition form factor in the same approximation:
\begin{eqnarray}
 Q^2 F_{\gamma^*\gamma\to \pi}(Q^2) &=& \frac{\sqrt{2}f_\pi}{3}\biggl\{
 \int\limits_0^1 \frac{dx}{x} \phi_\pi(x)+ \int\limits_0^{x_0} dx
\left[\frac{Q^2}{\bar x m^2_\rho} e^{(\bar x m^2_\rho-xQ^2)/(\bar x M^2)} - \frac{1}{x}\right]\phi_\pi(x)
\biggr\}
\nonumber\\&\equiv&  \frac{\sqrt{2}f_\pi}{3}\left(\int\limits_0^1 \frac{dx}{x} \phi_\pi(x)\right)
\biggl[1+ \frac{\widehat\xi_{\gamma^*\gamma\to\pi}(Q^2)}{Q^2}\biggr], 
\label{pionsoft}
\end{eqnarray}
where $x_0 = s_0/(s_0+Q^2)$. The results are shown for two sample models of the pion DA:  
1) the asymptotic DA $\phi_\pi(x)=6 x(1-x)$ and 2) a realistic  DA with nonvanishing 
Gegenbauer moments $a_2( 1 \mbox{GeV}) = 0.17$ and $a_4( 1 \mbox{GeV}) =0.06$.  
These are typical numbers that are used in the light-cone sum rule analysis  
of the pion electromagnetic  form factor \cite{pionffLCSR} 
and weak $B\to\pi\ell \nu_\ell$ decays \cite{KMOW},
see also a discussion in \cite{Agaev:2010aq}.   

Note that the size of the soft correction to the $\gamma^*\gamma\to \pi$ form factor 
for the asymptotic pion DA is very similar to the soft correction to $B\to \gamma \ell \nu_\ell$
for the existing models (and for $\lambda_B= 500$~MeV). For the realistic pion DA the correction
is larger. The difference is partially due to the larger value of the inverse moment
$\int\limits (dx/x) \phi_\pi(x)= 3(1+a_2+a_4+\ldots)$ playing the same role as $\lambda^{-1}_B$ for $B$-meson,
but also due to the functional form.
If the pion DA has some enhancement close to the end points, as was suspected in particular in view of the 
BABAR data \cite{BABAR} but also not excluded by Belle \cite{BELLE}, 
then the soft correction is larger and the onset of the asymptotic regime 
(where it is a $1/Q^4$ correction) occurs much later, see \cite{Agaev:2010aq,BMPS}.

Varying the inverse moment of the $B$-meson DA 
in the interval $300\, \text{MeV} < \lambda_B < 600 $~MeV 
we obtain the following values of the rescaled soft factor $\widehat{\xi}$, defined in (\ref{hatxi}),
averaged over the photon energy interval  $2\,\text{GeV}\,< 2 E_\gamma < m_B$:\\[1.5mm]
%
%\begin{table}[ht]
\renewcommand{\arraystretch}{1.2}
\begin{center}
\begin{tabular}{|c|c|c|c|c|} \hline
 $\lambda_B$ [MeV]       & 300     & 400    & 500     & 600    \\ \hline 
$\widehat{\xi}_{Model \,I}$ [GeV] & $-\left(0.50^{+0.04}_{-0.12}\right)$ & $-\left(0.36^{+0.06}_{-0.11}\right)$ & $-\left(0.27^{+0.07}_{-0.09}\right)$ & $-\left(0.22^{+0.07}_{-0.08}\right)$ \\ \hline 
$\widehat{\xi}_{Model \,II}$ [GeV] &     &          & $-\left(0.23^{+0.08}_{-0.09}\right)$ &  \\ \hline 
\end{tabular}
\end{center}
\renewcommand{\arraystretch}{1.0}
%\end{table}
\vspace*{3mm}\noindent
The quoted uncertainties include variations of the Borel mass 
and the photon energy dependence. 

Our result for $\widehat{\xi}$ (Model I) translates to the value of the 
coefficient $c$ in the notation of (\ref{eq:cxi}):
\begin{equation}
  c = \left(\frac{m_B}{2 E_\gamma}\right)\frac{e_u}{\lambda_B} \widehat{\xi}^{(0)}
    =\left(\frac{m_B}{2 E_\gamma}\right)\times\left\{ \begin{array}{ll}
 -(1.11^{+0.09}_{-0.27})\,,&\lambda_B=300\, {\rm MeV}\\[1mm]
   -(0.60^{+0.10}_{-0.18})\,,             &\lambda_B=400\,{\rm MeV}\\[1mm]
 -(0.36^{+0.09}_{-0.12})\,, &\lambda_B=500 \,{\rm MeV}\\[1mm]
 -(0.24^{+0.08}_{-0.09})\,,&\lambda_B=600\,{\rm MeV}
\end{array}
\right.\,,
%-(0.36\pm 0.12)\left(\frac{500\,\text{MeV}}{\lambda_B}\right)^{2.2\pm 0.5} \frac{m_B}{2 E_\gamma}\,.
\label{eq:cnum}
\end{equation}
where the uncertainties  have the same origin as in the above Table.
We emphasize that the soft contribution $\xi(E_\gamma)$ obtained here has
a suppression factor $1/(2E_\gamma)$ with respect to the leading term, as
compared to the $1/m_B$ suppression assumed in \cite{Beneke:2011nf}.
Note that for large $\lambda^{-1}_B$, i.e. for the $B$-meson DA that is enhanced in the 
soft region, there is a strong cancellation between the leading term and the soft
contribution: In a hypothetical limit $\lambda_B \to 0$ both terms diverge but the sum of them remains finite. 

%%%%%%%%%%%%%%%%%%%%%%%%%
%%%%%%%%%%%%%%%
\vspace*{0.3cm}

\begin{figure}[t]
\psfrag{x}[cc][cc][0.9]{$E_\gamma\, \text{[GeV]}$}
\psfrag{a}[cc][cc][0.9]{$\widetilde{F}_{A}(E_\gamma)$}
\psfrag{v}[cc][cc][0.9]{$F_{V}(E_\gamma)$}
\psfrag{i}[cc][cc][0.9]{$\xi(E_\gamma)$}
\begin{center}
\includegraphics[width=.45\textwidth,clip=true]{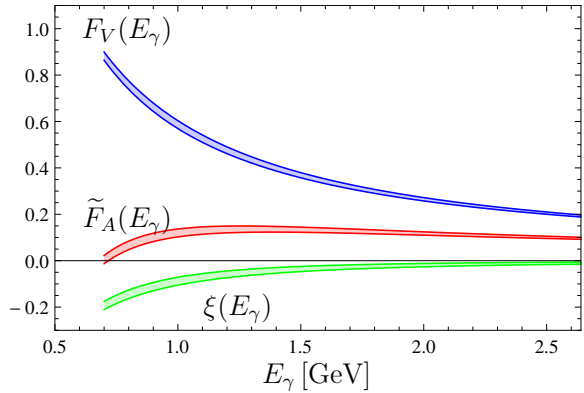}\hspace*{0.3cm}
\end{center}
\caption{\small\sf                                                
The vector and axial form factors for $B\to \gamma \ell\nu_\ell$  
calculated from  (\ref{eq:SCET1}) 
at $\lambda_B=500$ MeV. The lowest curves show the 
soft-overlap part $\xi(E_\gamma)$ of the form factors. 
The shaded areas indicate the  uncertainties
estimated by varying the Borel parameter  within 
$M^2=1.0- 1.5 $~GeV$^2$ and switching from the model I to model II 
of the $B$-meson DA.}
\label{fig:FAFV}
\end{figure}

{\large \bf 5.} 
We can employ our estimate of the soft-overlap contribution 
to calculate the  form factors $F_V$  and $F_A$  
at  large photon  energies. To this end we use the expressions in (\ref{eq:SCET1})
and the result for $R(E_\gamma,\mu)$ given in \cite{Beneke:2011nf}
which includes resummation of radiative corrections to the next-to-leading order 
logarithmic accuracy. Following \cite{Beneke:2011nf}
we adopt the ``soft-collinear'' scale  $\mu=1.5 $ GeV, the 
heavy-quark mass scale  $m_b=4.8$ GeV and $f_B=195$ MeV.
In the context of this study we are interested mostly in the uncertainty due to the soft contribution;
the variation of the $B$-meson decay constant 
within the intervals of the current lattice QCD estimates, 
will yield an  additional (correlated) theoretical uncertainty  in both form factors
about $\pm 10 \%$.
The results for the  vector and axial form factors --- the latter including the contact
term ---  and, separately, for the soft contribution  defined in our
approximation as $\xi(E_\gamma)=\frac{e_uf_Bm_B}{\lambda_B(2E_\gamma)^2}
\widehat{\xi}^{(0)}(E_\gamma)$, are shown for the choice $\lambda_B= 500$ MeV in Fig.\,\ref{fig:FAFV}. 
From (\ref{eq:cnum}) it is clear that our estimate of the soft form factor has 
a smaller error than the interval $-1<c<1$ taken in \cite{Beneke:2011nf}.

\begin{figure}[t]
\psfrag{x}[cc][cc][0.9]{$\lambda_B\text{[GeV]}$}
\psfrag{y}[cc][cc][0.9]{$BR(B\to \gamma\ell\nu_\ell)_{E_\gamma>E_{min}}\times 10^6$}
\begin{center}
\includegraphics[width=.45\textwidth,clip=true]{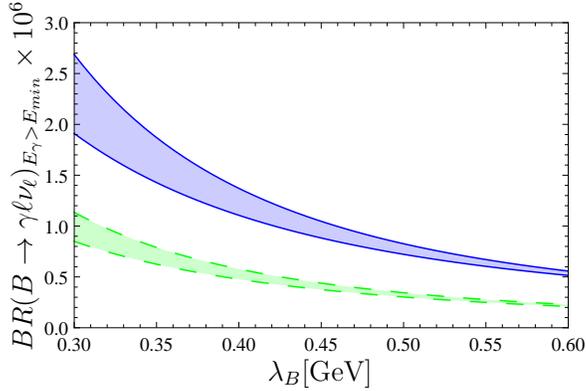}\hspace*{0.3cm}
\end{center}
\caption{\small\sf The partial branching fraction 
$BR(B\to \gamma\ell\nu_\ell)_{E_\gamma>E_{min}}$ for $E_{min}=1.0$ GeV
(upper, solid) and $E_{min}=1.7$ GeV (lower,dashed). 
The uncertainties indicated by shaded areas are of the same origin as in 
Fig.~\ref{fig:FAFV}.}
\label{fig:BR}
\end{figure}

Using these form factors we calculate the partial branching fraction $BR(B\to \gamma\ell\nu_\ell)$, 
integrating (\ref{eq:width}) over the photon energies 
$E_{min}<E_\gamma <m_B/2$ . The result is shown in Fig.~\ref{fig:BR}  as a function of 
$\lambda_B$ for two different choices of the photon energy cut,
$E_{min}=1.0\, (1.7) $~GeV.  For this plot, for definiteness, we take  $|V_{ub}|=3.5 \times 10^{-3}$,
in the ballpark of current determinations from exclusive semileptonic $B$ decays. 
%%%%%%%%%%%%%
%%%%%%%% 
Our main message is that the uncertainty due to the soft overlap contribution
is sufficiently small.

The existing experimental data are not yet conclusive.
The upper limit on the partial branching fraction 
$BR(B\to \gamma \ell \nu_\ell)$ 
with certain kinematical cuts, including the cut on $E_\gamma$,   
was   published by the BABAR collaboration in \cite{Aubert:2009ya}. 
This limit  is weaker than their previous result quoted in \cite{Aubert:2007yh}. 
As explained in \cite{Beneke:2011nf},  the published limit \cite{Aubert:2009ya} is not yet 
sufficient to constrain the inverse moment $\lambda_B$. 

Finally, we suggest to consider the ratio of the photoleptonic 
and leptonic charged $B$-meson decay widths:
\begin{eqnarray}
R_{\gamma\mu\nu/\tau\nu}(E_{min})&\equiv& \frac{ BR(B\to \gamma\mu\nu_\ell)_{E_\gamma>E_{min}}}{BR(B\to \tau \nu_\tau)}
\nonumber \\
&=&\!\!\!\frac{4\alpha_{em}}{3\pi m_\tau^2(1-m_\tau^2/m_B^2)^2}
\!\!\int\limits_{E_{min}}^{m_B/2} \!\!\! \!dE_\gamma
E_\gamma^3\!\left ( 1 -\frac{2E_\gamma}{m_B} \right) 
\Big[ \frac{|F_V(E_\gamma)|^2+|\widetilde{F}_A(E_\gamma)|^2}{f_B^2} \Big], 
\label{eq:ratio}
\end{eqnarray}
where we neglect the muon mass. 
Both observables are accessible in the $B$-factory experiments and their ratio
does not depend on $V_{ub}$ and on the $B$-meson decay constant because 
$f_B$ enters  also  the normalization of the form factors.
We predict: $R_{\gamma\mu\nu/\tau\nu}(E_{min}=1.7$~GeV$)=0.0103, 0.0058,0.0037,0.0025$ 
at $\lambda_B=300,400,500,600$ MeV, respectively (for model I, $M^2$ =1.0 GeV$^2$). 
Note that the recent measurement 
\cite{BelleBtaunu}  of $B\to \tau \nu_\tau$ yields  
$f_B|V_{ub}|=[7.4\pm 0.8(stat)\pm0.5(syst)]\times10^{-4}$~GeV, 
consistent with the lattice-QCD value of $f_B$ 
and with the value of  $V_{ub}$  which we have used above 
for the estimate of the partial photoleptonic $B$ decay width.

 %%%%%%%%%%%%%%%%%%%5
%%%%%%%%%%%%%%%%%
%%%%%%%%%%%%%%%%%%%%%%%%%%%%%%%%%%%%%%%%%%%%%%%%%%%%%%%%%%
%%%%%%%%%%%%%%%%%%%%%%%%%%%%%%%%%%%%%%%%%%%%%%%%%%%%%%%%%%

\vspace*{0.3cm}

{\large \bf 6.}~~
To conclude, we have described a method to calculate the soft contribution to
the decay $B\to \gamma \ell \nu_\ell$ which is formally subleading in powers of 
the photon energy $E_\gamma$ and is not directly accessible in QCD factorization.
This method has originally been developed for another process, 
the $\gamma^* \gamma \to \pi$ transition form factor, in which case the 
QCD calculation is under a better control because the moments of pion DA can be
calculated in lattice QCD. The successful description of the experimental 
data on the $\gamma^* \gamma \to \pi$ form factor in this framework 
in the region of momentum transfers 
$Q^2 =2-6$~GeV$^2$  (see e.g.~\cite{Agaev:2010aq} and references therein) 
allows one to hope that the same technique will yield sufficiently accurate 
predictions for the decay $B\to \gamma \ell \nu_\ell$ as well.

The calculation presented in this letter serves the purpose of demonstration mainly. 
It can and should be extended in several directions, by replacing 
the leading-order expression for the spectral density 
$\mathrm{Im} F^{\rm HQE}_{B\to\gamma^*}$ in (\ref{eq:general}) by the complete 
HQE expression to the $\mathcal{O}(1/(2E\gamma), 1/m_B$) accuracy.  
First of all, one has to include radiative corrections which give rise 
to the $R(E_\gamma,\mu)$ factor in (\ref{eq:SCET1}).
This requires a calculation of the terms $\mathcal{O}(\alpha_s)$ in 
the coefficient function of the $B$-meson DA for nonzero photon virtualities,
which is straightforward. Such corrections can then be resummed using SCET 
techniques, although we expect that the numerical impact of the 
resummation will be minimal.

Second, one has to include contributions of higher-twist two-particle and also 
three-particle $B$-meson DAs. These contributions contain logarithmic 
end-point singularities if calculated directly, but give rise to finite contributions 
to the dispersion integral (\ref{eq:general}) with the continuum threshold 
$s_0$ providing an effective IR cutoff. In other words, the full contribution 
of twist-four DAs to the form factor is finite, but an attempt to rewrite the answer
as a sum of the ``pure'' HQE expression plus a correction, as in (\ref{eq:LCSR_LO1})
would result in divergent expressions. We believe that the corresponding calculation
would be very interesting because the normalization constants in the higher-twist
$B$-meson DAs are related to higher moments of the leading-twist DA, 
at least within certain schemes used to subtract the corresponding high-energy behavior,
see \cite{Grozin:1996pq,Lee:2005gza,Kawamura:2008vq,Nishikawa:2011qk}.

Third, there are terms related to photon emission at large distances
that involve a photon DA. These contributions were estimated for the 
$\gamma ^*\gamma \to \pi$ transition form factor in \cite{Agaev:2010aq}
in which case they correspond to contributions of {\em twist-six} operators
in the operator product expansion. This example is instructive as it shows that 
for soft corrections the correspondence between power suppression and 
twist counting is lost: Contributions of {\em all}\ twists to the operator product
expansion of the electromagnetic form factors produce power corrections which 
are suppressed by {\em one} power of $Q^2$ with respect to the leading twist result.
We expect that the situation with the HQE for the decay $B\to \gamma \ell \nu_\ell$
is similar. 

Last but not least,  let us mention that 
$B\to \gamma\ell\nu_\ell $ decay amplitudes at large photon energies have also been calculated
using light-cone sum rules with photon DA's and $B$-meson interpolating current 
\cite{Khodjamirian:1995uc,Eilam:1995zv,Ball:2003fq}.
It would be interesting to clarify the interconnection of this approach with 
the methods employing the $B$-meson DA's in order to gain a more complete picture of 
the underlying quark-gluon dynamics.

%%%%%%%%%%%%%%%%%%%%%%%%%%%%%%%%%%%%%%%%%%%%%%%%%%%%%%%%%%
%%%%%%%%%%%%%%%%%%%%%%%%%%%%%%%%%%%%%%%%%%%%%%%%%%%%%%%%%%
%%%%%%%%%%%%%%%%%%%%%%%%%%%%%%%%%%%%%%%%%%%%%%%%%%%%%%%%%%

\newpage

\end{document}